# Beyond Superexchange: Emergent Unconventional Ferromagnetism in Thin-Film Sandwich Structures of Intrinsic Magnetic Topological Insulators

Takuya Takashiro[1,*,†], Ryota Akiyama[1,‡], Ryotaro Minakawa[1], and Shuji Hasegawa[1]

[1]*Department of Physics, The University of Tokyo, Bunkyo, Toyko 113-0033, Japan*

**Abstract**

An intrinsic magnetic topological insulator $Mn(Bi_{1-x}Sb_x)_2Te_4$ (MBST in short), possesses highly-ordered ferromagnetic Mn-atom monolayer self-assembled at the center of septuple-layer (SL) stacking unit, and thus is expected to provide a promising platform for quantum phenomena such as quantum anomalous Hall effect (QAHE) at high temperatures. We investigate the nature of magnetic interactions in MBST systems by fabricating two types of structures: $Mn(Bi_{1-x}Sb_x)_2Te_4$(1 SL)/$(Bi_{1-x}Sb_x)_2Te_3$ heterostructure (MBST/BST in short) having only intralayer magnetic interaction in single MBST layer, and MBST(1 SL)/BST/MBST(1 SL) sandwich structure having the interlayer magnetic interaction between the two MBST layers, in addition to the intralayer interaction. The out-of-plane magnetization in the sandwich structure turned out to be significantly larger than that in the heterostructure, indicating that the interlayer interaction between two

[*] Present address: Institute of Semiconductor and Solid State Physics, Johannes Kepler University, Linz, 4040, Austria
[†] takuya.takashiro@jku.at
[‡] akiyama@surface.phys.s.u-tokyo.ac.jp

MBST layers is ferromagnetic though it had been described as antiferromagnetic in previous research. Additionally, by partially substituting Bi by Sb atoms to change the in-plane lattice constant in MBST layer, the Curie temperature ($T_C$) increased with a smaller in-plane lattice constant. That is, the intralayer ferromagnetic interaction within each MBST layer is enhanced by decreasing the in-plane Mn-Mn distance. These behaviors cannot be explained only by the direct exchange nor superexchange interactions reported previously for MBST systems. Particularly, in the sandwich structure with $x = 0$ (without Sb), $T_C$ in the case of $Bi_2Te_3$-spacer of 2 – 9 quintuple-layer thick was higher than that in the case without the spacer layers, which indicates that the increase of spacer thickness changes the sandwich structure from trivial to nontrivial topological states. This enhancement of ferromagnetism is proposed to be explained by the van Vleck mechanism. Furthermore, the curve shape of anomalous Hall effects for the sandwich structure with $x = 0.55$ was almost unchanged by carrier-density modulation with applying the top gate voltage, which suggests that the ferromagnetism in this system is carrier-independent, reinforcing the above-mentioned mechanism. This work provides a new perspective on magnetic mechanisms in MBST systems, which helps us to realize next-generation spintronic and electronic devices by flexibly controlling magnetism in intrinsic magnetic topological insulators.

**I. INTRODUCTION**

Introducing magnetic order into a topological insulator (TI) results in breaking down the time-reversal symmetry, leading to the gapped Dirac-type surface state [1,2,3]. This interplay between magnetism and topology gives rise to exotic quantum phenomena such as quantum anomalous Hall effect (QAHE) [4-8], chiral Majorana modes [9], and magnetic skyrmions [10-14], which are promising for application to dissipation-less magnetic memory, quantum computers, and so on. So far, most attempts to realize magnetic TIs have been performed by random and dilute doping of magnetic atoms into TI crystals [5,6,7]. However, this approach causes inhomogeneous magnetic interactions in systems and thus the observation of QAHE is limited to extremely low temperatures (0.03-2 K) [7]. In order to increase the observation temperature of such quantum phenomena ($T_Q$) for application, it is required that magnetic atoms are highly ordered in TI crystals.

Recently, an intrinsic magnetic topological insulator (IMTI) compound, Mn(Bi$_{1-x}$Sb$_x$)$_2$Te$_4$ (MBST in short), has attracted much attention as an alternative to the diluted-magnetic TIs [8,14-42]. MBST consists of a septuple layer (SL) as the unit of atomic layer stacking where one Mn-Te layer is incorporated at the center of each quintuple layer (QL) of a TI crystal (Bi$_{1-x}$Sb$_x$)$_2$Te$_3$ (BST in short), which leads to realizing the self-assembly highly-ordered ferromagnetic single layer of Mn atoms as illustrated in Fig. 1(a) (generated with VESTA [43]). Starting with our report of the prototype of self-assembled IMTIs, Se-based MnBi$_2$Se$_4$, grown on Bi$_2$Se$_3$ layers in 2017 [44], vigorous types of novel quantum phenomena such as skyrmions in 2022 [14] and QAHE in 2020 [8] were reported in epitaxial films and bulk single crystals composed of Te-based MBST and BST. Nevertheless, although such observations of the emergence of ferromagnetism and accompanying phenomena have been reported, much remains unknown about the nature of ferromagnetism itself, so that seemingly controversial results have been reported. It is known that spins of Mn within each SL of MBST are ferromagnetically coupled with each other due to the intralayer interaction while adjacent SLs are known to be antiferromagnetically coupled by the interlayer interaction [8,16,18,21-30,32,33]. On the other hand, we previously observed ferromagnetic behavior in a sandwich structure where two SLs of MBST were separated by nonmagnetic BST layers of 0 – 2 QL thick [14]. This is inconsistent with those of multi-layered structures like bulk single crystals of MBST to be antiferromagnetic [18,22-24,26-30,32,33]. Besides, it was reported that the Curie temperature ($T_C$) of one SL of $MnBi_2Te_4$ is as high as ~ 15 K because the superexchange interaction among Mn atoms within this layer [21,30,32,33] mediates ferromagnetism whereas a $MnBi_2Te_4/Bi_2Te_3$ heterostructure (one SL formed on the $Bi_2Te_3$ layer) was reported to be paramagnetic even at 6 K [34]. Meanwhile, as a mechanism for the magnetic interaction, Padmanabhan *et al.* experimentally verified that magnetization in the $MnBi_2Te_4$ bulk crystal is not induced by the superexchange model, but by the *p-d* exchange coupling between localized spins and topological-band electrons [41].

Therefore, it is strongly required now to reconcile these discrepancies and unveil the magnetic mechanism for realizing high $T_Q$ of IMTI systems.

In this paper, to reveal the whole picture of magnetism in MBST systems, we elucidate the magnetic interactions in two types of model samples as illustrated in Fig. 1(b): the MBST(1-SL thick)/BST heterostructure having only intralayer magnetic interaction, and the MBST(1-SL thick)/BST($N$-QL thick)/MBST(1-SL thick) sandwich structure ($N$=0-9) having both the intra- and interlayer magnetic interactions. From measurements of the magnetization and anomalous Hall effect (AHE), it was found that the magnetization amplitude and $T_C$ of the sandwich structure were significantly larger than those of the heterostructure. This means that the interlayer interaction between two MBST layers is ferromagnetic. Then, the spacer-thickness-dependence of ferromagnetism in the sandwich structure suggests the long-range coupling between two MBST layers. On the other hand, Sb-content ($x$)-dependences of $T_C$ in both structures demonstrated that the intralayer ferromagnetic interaction was enhanced by decreasing the Mn-Mn distance within each MBST layer via the increase of $x$ in each MBST layer. These results do not agree with the direct/superexchange interactions which were previously proposed as magnetic mechanisms in MBST bulk crystals [18,22-24,26-30,32,33]. Besides, the carrier-density modulation by changing the top (ion liquid) gate voltage did not change AHE properties in the sandwich structure sample, which implies that the ferromagnetic interaction is carrier-independent. Based on these experimental results, we propose a ferromagnetic model realized in the present systems.

## II. EXPERIMENTAL DETAILS

The samples of heterostructure and sandwich structure were grown on *n*-type Si(111) substrates by molecular beam epitaxy (MBE) technique in an ultrahigh vacuum chamber equipped with a reflection high-energy electron diffraction (RHEED) system. The MBST-SL was self-assembly formed by depositing one layer of MnTe on top of the BST layer at ~ 270 ºC as described in Supplemental Material (SM) [45] and our previous report [14]. Note that separate sources of Mn and Te were used during the deposition, which was different from the method with a compound source of MnTe in a previous report [42]. Then, we deposited Al on top of the surface of the hetero- and sandwich structures at room temperature to form a ~ 2-nm-thick $Al_2O_3$ cap layer after taking it out into air.

In addition to the RHEED observation *in situ* during the growth, the crystal quality and lattice constant of thin films were evaluated *ex situ* by X-ray diffraction (XRD) system of SmartLab (RIGAKU Corp.). The layers of BST and MBST are epitaxially grown in their (001) crystallographic orientation with the hexagonal unit cell on Si(111) substrates. To estimate the in-plane lattice constants $a$ ($b$), we measured the 015 diffraction of BST lattice by inclining the sample surface with respect to X-ray beam, in addition to the 001 diffraction for estimating the out-of-plane lattice constant $c$, because the 015 peak position contains information about both out-of-plane and in-plane lattice constants (see SM for more details) [45].

Magnetic measurements were performed using the superconducting quantum interference device (SQUID) at Magnetic Property Measurement System (MPMS3, Quantum Design, Inc.). To extract the magnetization of the grown layers, the diamagnetic component of the sample from e.g., the substrate was subtracted from raw data, which was estimated from the linear component in the observed magnetization curves as a function of the applied magnetic field ($M$-$H$ curves) at high magnetic field region (4-5 T) at 2 K.

Electrical transport measurements were performed using the conventional six-terminal method in Physical Property Measurement System (PPMS, Quantum Design, Inc.). Au wires were bonded on the sample with silver paste to apply an in-plane electrical current (up to 10 $\mu$A). Magnetic field (up to 14 T) was applied normal to the sample surface (along $c$ axis). To tune the Fermi level ($E_F$) effectively, an electric double-layer capacitor containing an ionic liquid was used for gating. Note that the ionic liquid, *N, N*-diethyl-*N*-(2-methoxyethyl)-*N*-methylammonium bis-(trifluoromethylsulfonyl) imide (DEME-TFSI)), was dissolved by the polymer, poly(styrene)-b-poly(ethylene) oxide-b-poly(styrene), using acetone and then the solution was dropped on the $Al_2O_3$-capping layer on the samples [46,47].

## III. EXPERIMENTAL RESULTS

### A. Crystal structure characterization

Figure 1(c) displays XRD $\theta$-$2\theta$ scan profiles of the MBST(1 SL)/BST(8 QL) heterostructures with various Sb contents ($x$) grown on Si(111) substrates. These diffraction peaks indicate that the thin film is single-crystalline and epitaxially grown in the (001) crystallographic orientation on the substrate. In addition, one can note that the 009 and 0012 peaks, which are forbidden reflections for $Bi_2Te_3$ crystal, appear and are enhanced with increasing $x$. These have also been reported in BST thin film systems, which can be interpreted by breakdown of extinction law in XRD due to a change of the structure factor caused by the substitution of Sb atoms for Bi atoms in BST [48].

Figure 2(a) displays 015 diffraction peaks of BST lattice in the MBST/BST heterostructures ($x$ = 0, 0.3, 0.55, 0.8, and 1, respectively), showing peak shifts with increase of *x*. The in-plane lattice constant *a* (*b*) can be estimated from the 015 diffraction peak using the lattice parameter *c* obtained from Fig. 1(c). Figures 2(b) and (c) show *x*-dependences of the in-plane lattice constant of the heterostructure and sandwich structure, respectively. In both systems the in-plane lattice constant tends to decrease with increasing *x*. This is because of the lattice distortion due to partial substitution of Bi atoms by smaller Sb atoms in MnBi$_2$Te$_4$ as reported by bulk MBST systems [23].

**B. Interlayer magnetic interaction**

To investigate the interlayer magnetic interaction between two MBST layers, we compare magnetic properties of the MnBi$_2$Te$_4$(1 SL)/Bi$_2$Te$_3$(2 QL)/MnBi$_2$Te$_4$(1 SL) sandwich structure with those of the MnBi$_2$Te$_4$(1 SL)/Bi$_2$Te$_3$(8 QL) heterostructure (both are *x* = 0 samples), by magnetization measurements with MPMS3. Note that the applied magnetic field was normal to the film surface. Figures 3(a) and (b) display *M-H* curves at *T* = 2 K and temperature dependences of magnetization (*M-T*) under the magnetic field of *H* = 400 Oe, respectively. The magnitude of magnetization here is normalized for 1 SL ($M_{1SL}$). The blue and red curves indicate the properties of the heterostructure and sandwich structure, respectively. From two *M-H* curves in (a), the remanent magnetization of the sandwich structure is ~18 times as large as that of the heterostructure. As shown in two *M-T* curves in (b), the magnetization enhancement with decreasing temperature in the sandwich structure is ~7 times as large as that in the heterostructure. Both results indicate that the ferromagnetic order is enforced by the interlayer interaction in the sandwich structure. Intriguingly, there are no signs of antiferromagnetism that would show up as *M-H* steps and *M-T* kinks as previously reported in multi-layered systems [16,18,22-25,27,29,30]. This suggests that the interlayer ferromagnetic interaction between the top and bottom SLs of MnBi$_2$Te$_4$ in the sandwich structure is different from that in the multi-layered (bulk) system.

Next, we investigate the interlayer interaction in the MnBi$_2$Te$_4$/Bi$_2$Te$_3$/MnBi$_2$Te$_4$ sandwich structure by varying the thickness of the Bi$_2$Te$_3$ spacer layer which corresponds to the distance between the top and bottom ferromagnetic MnBe$_2$Te$_4$ layers. Figures 4(a) - (d) display *M-H* curves at 2 K of the sandwich structures having the Bi$_2$Te$_3$ spacer-thickness (*N* QL) with *N* = 0, 2, 6, and 9, respectively. *M-H* curves with every *N* show ferromagnetic hysteresis loops with the coercivity ($H_C$) of ~ 1.2 kOe. More importantly, the remanent magnetization normalized per 1 SL in the *N* = 9 sample is ~ 15 times as large as that of the MnBi$_2$Te$_4$/Bi$_2$Te$_3$ heterostructure having no interlayer interaction shown in Fig. 3(a). This indicates that the ferromagnetic coupling between the top and bottom SLs works even for the spacer thickness of 9 QL.

Figure 5 shows how $T_C$ in the sandwich structure changes depending on the spacer layer thickness. Figure 5(a) displays *M-H* curves in the MnBi$_2$Te$_4$/Bi$_2$Te$_3$/MnBi$_2$Te$_4$ sandwich structures with $N = 0$ and 2 at temperatures ranging from 2 to 20 K under the magnetic fields perpendicular to the plane, respectively. As shown in Fig. 5(b), $T_C$ of samples with $N = 0$ and 2 are estimated to be 13.8 $\pm$ 0.6 K and 18.7 $\pm$ 0.5 K by Arrott plots of the magnetization in (a), respectively. Both $T_C$ are larger than that of the MnBi$_2$Te$_4$/Bi$_2$Te$_3$ heterostructure ($T_C$ = 6.6 $\pm$ 0.2 K) as mentioned in the next section. This enhancement of $T_C$ is due to the ferromagnetic interlayer interaction in the sandwich structure. Figure 5(c) represents the dependence of $T_C$ on the spacer layer thickness in the MnBi$_2$Te$_4$/Bi$_2$Te$_3$($N$ = 0 – 9)/MnBi$_2$Te$_4$ sandwich structure. Remarkably, $T_C$ in the sample without the spacer ($N = 0$) is lower ($T_C$ = 13.8 $\pm$ 0.6 K) than that of samples with $N = 2 – 9$ ($T_C$ = 17.5 $\pm$ 0.5 K - 18.7 $\pm$ 0.5 K). This is because, as discussed later in Section IV C, too strong hybridization between the top and bottom surface states (finite size effects) makes the $N = 0$ system topologically trivial, leading to the suppression of band-inversion-induced ferromagnetism.

**C. Intralayer magnetic interaction**

Let us now discuss on the intralayer magnetic interaction within each SL. We systematically evaluated magnetic properties of MBST/BST(8 QL) heterostructures and MBST/BST(6 QL)/MBST sandwich structures with various *x* (while all MBST layers are 1 SL thick). Here, this spacer-thickness 6 QL of sandwich structures is adopted because it is thick enough to suppress the hybridization between the top and bottom surface states [14,39]. Figure 6(a) shows *M-H* curves in MBST/BST heterostructures with $x$ = 0, 0.3, 0.8, and 1 at several temperatures within 2 K – 10 K. Figure 6(b) represents Arrott plots corresponding to curves in Fig. 6(a). In the case of $x$ = 0, the left panel in Fig. 6(b), MnBi$_2$Te$_4$/Bi$_2$Te$_3$, $T_C$ = 6.6 ± 0.2 K is estimated, which is comparable to that of MnBi$_2$Te$_4$/Bi$_2$Te$_3$ in XMCD measurements reported by Fukasawa, *et al.* ($T_C \leq$ 5.6 K) [34].

Figure 7(a) summarizes the $x$-dependence of $T_C$ in the MBST/BST heterostructure given by Arrott plots of magnetization in Fig. 6(b). Intriguingly, $T_C$ rises by increasing $x$, meaning that the ferromagnetic interaction becomes stronger. The enhanced ferromagnetic interaction here is intralayer only, because the system has only one magnetic MBST layer. Figure 7(b) shows the $x$-dependence of $T_C$ (estimated by Arrott plots of magnetization) in the MBST/BST/MBST sandwich structure where both the intra- and interlayer exchange interactions can work. As in the heterostructure case, $T_C$ in the sandwich structure is more enhanced with increasing $x$. Here, note that by increasing $x$, the hole density also increases and the carrier type switches from $n$- to $p$-types around $x$ = 0.7 in the MBST/BST case (see SM section III [45]) and around $x$ = 0.55 in the MBST/BST/MBST case (see the work we previously reported [14]), respectively. It has been reported for Mn-doped $Bi_2Te_{3-y}Se_y$ that if Dirac carriers mediate ferromagnetic ordering, $T_C$ becomes larger as $E_F$ approaches the charge neutral point (CNP) [49]. As shown in Fig. 7, however, $T_C$ of both MBST/BST and MBST/BST/MBST cases does not take maximum in $T_C$ when $x$ is around the Sb-content for CNP ($x$=0.7 and 0.55 for the two cases, respectively), but rather tends to monotonically increase by increasing $x$, which is clearly different from that of the Dirac-carrier-mediated ferromagnetic system.

**D. Carrier density dependence of the magnetic property**

As mentioned in the previous section, an increase of $x$ leads to a decrease in the in-plane lattice constant and an increase in the hole density (that is, shifting $E_F$ downward, see SM). To separate these two effects, we performed Hall measurements for the sandwich structure with changing the carrier density by ion-liquid gating method. With this, we can investigate the effect on magnetic properties by changing only the carrier density and keeping the lattice parameters unchanged. Figure 8(a) displays magnetic field dependences of the Hall resistivity ($\rho_{yx} - \mu_0 H$) in the MBST/BST(6 QL)/MBST sandwich structure with $x = 0.55$ at $T = 2$ K at various top gate voltages ($V_g$) ranging from -2.5 V to + 1.5 V. $\rho_{yx}$ in a ferromagnetic material is described in general as $\rho_{yx} = R_H \mu_0 H + \rho_A$, where the first term denotes the ordinary Hall resistivity with the ordinary Hall coefficient $R_H$ which depends on the carrier type/density, and the second term is the anomalous Hall resistivity $\rho_A$ which features ferromagnetic properties. Figure 8(b) shows magnetic field dependences of the ordinary Hall resistivity estimated from the linear slope of $\rho_{yx}-\mu_0H$ at the region of high magnetic field in Fig. 8(a). The slope of the ordinary Hall resistivity, namely $R_H$, changes from positive to negative by varying $V_g$ from negative to positive, which indicates that the carrier type and density are modulated systematically by ion-gating.

Figure 9(a) depicts $V_g$-dependences of the ordinary Hall coefficient $R_H$ (top) and the coercivity $H_C$ (bottom) in the sandwich structure obtained from the Hall resistivity shown in Fig. 8(a). To compare with the effect of the Sb-substitution, $x$-dependences of $R_H$ (top) and $H_C$ (bottom) with changing $x$ in the sandwich structure are displayed in Fig. 9(b). From the top panel of Fig. 9(a), the sign of $R_H$ switches near $V_g$ = -1.5 V, which corresponds to CNP (judged from Fig. 8(b)). $R_H$ is near zero with the top gate voltage going close to the CNP voltage. This is because both electrons and holes contribute to the transport due to $E_F$ close to CNP [50, 51]. This situation achieved by the gating corresponds to the purple region near CNP in Fig. 9(b) (top) achieved by the Sb-substitution. Remarkably, in Fig. 9(a) (bottom) $H_C$ is almost constant by the gating despite the switch of the carrier type, whereas as shown in Fig. 9(b) (bottom) the change of $H_C$ by the Sb-substitution is more significant in the purple region near CNP. These results provide a suggestion that the change in $H_C$ by the Sb-substitution in Bi sites is due to the change in the lattice constant, rather than the carrier type/density, implying that the ferromagnetism is not carrier-mediated. Such carrier-independent mechanism was also reported for a Cr-doped BST system where $H_C$ was almost unchanged even by changing the carrier density [52].

## IV. DISCUSSION

### A. Magnetic interaction in MnBi$_2$Te$_4$ systems

MnBi$_2$Te$_4$ systems, *i.e.* $x = 0$ in Mn(Bi$_{1-x}$Sb$_x$)$_2$Te$_4$, including multi-layered thin films or bulk crystals have been reported by many groups so far, and the origins of magnetic interaction are explained by the direct/superexchange couplings [20,21,33]. Considering the interlayer magnetic interaction in bulk/multi-layered MnBi$_2$Te$_4$ crystals, one Mn atom within a SL is coupled to another one within the neighboring SL by the orbital overlap in the straight "chain" of Mn-Te-Bi-Te-Te-Bi-Te-Mn across the layers, which realizes antiferromagnetism between layers based on the superexchange given by the Goodenough-Kanamori-Anderson (GKA) rule [21, 53 - 55 ]. Furthermore, this interlayer interaction has been reported to disappear when the distance between neighboring SLs is increased by inserting 3 QL Bi$_2$Te$_3$ in-between [30]. However, in our samples as shown in Figs. 3 – 5, the interlayer interaction in the MnBi$_2$Te$_4$/Bi$_2$Te$_3$/MnBi$_2$Te$_4$ sandwich structure realizes long-range ferromagnetism and works even if two SLs are separated by ~ 9 QL Bi$_2$Te$_3$ from each other. These are inconsistent with the superexchange interlayer model which has been widely accepted in multi-layered MnBi$_2$Te$_4$ systems [21,33]. Additionally, we demonstrate that the saturated magnetization of 3 SL thick MnBi$_2$Te$_4$ is smaller than that of 2 SL one as displayed in Fig. S9 in SM, which implies the weaker ferromagnetic interaction of the 3 SL sample [45]. This implies that the mechanism of the interlayer interaction changes with increasing the number of stacking SLs, which may lead to the antiferromagnetism in such bulk or multi-layered systems.

On the other hand, the intralayer interaction within a SL is ferromagnetic due to the competition between the direct and superexchange couplings. It is known that while the direct exchange interaction between two Mn atoms prefers an antiferromagnetic order, the superexchange interaction mediated by neighboring Te $p$-orbitals induces ferromagnetic order according to the GKA rule [20,21,33,53-55]. In some theoretical and experimental reports assuming the competition between these interactions, $T_C$ of a single MnBi$_2$Te$_4$ layer has been reported to be 12 - 20 K [20,32,33]. In our study, however, $T_C$ of the heterostructure of 1 SL MnBi$_2$Te$_4$/8 QL Bi$_2$Te$_3$ is found to be ~ 6 K, which is rather consistent with the result reported by Fukasawa *et al.* [34]. Thus, the intralayer interaction as well as the interlayer interaction in our system do not agree with the direct/superexchange coupling model.

**B. Modulation of magnetic interaction by the Sb-substitution**

As displayed in Figs. 6 – 9, ferromagnetic properties of the MBST/BST heterostructure and MBST/BST/MBST sandwich structure are modulated by changing the Bi/Sb ratio. In bulk MBST systems, it is known that by increasing Sb-content $x$, Mn atoms occupy Sb sites in each SL more because the electronegativity and the ionic size of Sb are similar to Mn rather than those of Bi. Some reports say that randomly distributed anti-sites of Mn-Sb tend to turn the interlayer magnetic interaction from antiferromagnetic to ferromagnetic [35,37,38]. In our systems, however, ferromagnetic properties are innately observed in all heterostructures and sandwich structures with any Sb-contents including the $x = 0$ case. Moreover, $T_C$ of the MBST/BST heterostructure is raised by increasing $x$ despite the absence of interlayer ferromagnetic interactions in the heterostructure. This indicates that $Mn^{Sb}$ anti-site defects which can reinforce the interlayer interaction are unlikely to be the origin of the contribution to the ferromagnetism of our systems. For this reason, the Sb-substitution in MBST systems can rather change the strength of the intralayer interaction due to the change of the in-plane

Mn-Mn distance within a SL.

It has been theoretically and experimentally reported that if the intralayer interaction is a competition state between the direct exchange (antiferromagnetic) and superexchange (ferromagnetic) couplings, the antiferromagnetic direct exchange coupling becomes stronger as the in-plane Mn-Mn distance decreases by substituting Sb for Bi, which leads to the frustrated ferromagnetic order [20,33]. In our case, both $T_C$ of the heterostructure and sandwich structure become higher with decreasing the in-plane Mn-Mn distance by such Sb-substitution as shown in Fig. 10. This tendency cannot be explained only by the direct/superexchange model and our results request the other model for the stabilization of ferromagnetism with the Mn atoms closer to each other in a SL.

**C. Possible mechanism of ferromagnetism**

In many magnetic TIs, origins of the long-range ferromagnetic order are explained by models such as the carrier-mediated mechanism like the Ruderman-Kittel-Kasuya-Yosida (RKKY) interaction [49,56-62], the carrier-independent mechanism like the Bloembergen-Rowland (BR) interaction [57,60-65], the van Vleck-type magnetism [4,41,66-69], and the hybrid of RKKY and van Vleck models [52,70,71]. In the RKKY model, a ferromagnetic interaction among magnetic atoms is mediated by polarized free carriers, and the magnitude of ferromagnetism depends on the carrier density, namely the $E_F$ position of the magnetic TI [49,57,60,61]. In contrast, the BR interaction is regarded as an in-gap version of the RKKY interaction [60,61], which was originally suggested for explaining ferromagnetism in an insulator [63,64]. With $E_F$ being in the band gap, local moments in magnetic atoms are ferromagnetically aligned by virtually excited carriers (not real carriers) and the strength of the coupling depends on the band gap size, not on the carrier density [60]. Particularly, in expressions of the BR and RKKY interactions, the ferromagnetic interaction becomes larger as two magnetic atoms approach closer each other [60,61], which is consistent with the $T_C$ enhancement by decreasing the in-plane Mn-Mn distance in our MBST systems as displayed in Fig. 10. Finally, in the van Vleck mechanism, local magnetic moments are coupled via valence electrons in a topological insulator. When the band inversion between the bulk conduction and valence bands occurs due to the strong spin-orbit coupling, the spin susceptibility of valence electrons becomes large, which leads to stabilizing ferromagnetism without the assistance of free carriers [4,68]. In consequence, ferromagnetic properties in the van Vleck mechanism are independent of the carrier density [67].

As already mentioned in Fig. 5(c), in the MnBi$_2$Te$_4$/Bi$_2$Te$_3$/MnBi$_2$Te$_4$ sandwich structure without the spacer-Bi$_2$Te$_3$ ($N = 0$), whose $E_F$ crosses the bulk conduction band, $T_C$ is lower than those for the cases with $N = 2 - 9$. This suggests that the ferromagnetism of MnBi$_2$Te$_4$/Bi$_2$Te$_3$/MnBi$_2$Te$_4$ is not due to a simple sum of the two magnetic MnBi$_2$Te$_4$ layers. According to some reports on angle-resolved photoemission spectroscopy, MnBi$_2$Te$_4$/MnBi$_2$Te$_4$ ($N$ = 0) has no Dirac surface state [36] unlike that with the spacer layer Bi$_2$Te$_3$ ($N$ = 4) in-between [39]. Without a spacer layer, the inversion between the bulk conduction and valence bands does not occur because of finite-size effects, and the system becomes topologically trivial. In the van Vleck model, the ferromagnetic order is reported to be unstable when the bulk bands are not inverted [66,68]. Therefore, the spacer-thickness-dependence of $T_C$ in our system (Fig. 5(c)) reflects the inverted or non-inverted bulk bands, that is, the band topology. However, regarding ferromagnetism, it survives even in the case of $N$ = 0 although $T_C$ becomes lower. Because the $E_F$ position of MnBi$_2$Te$_4$/Bi$_2$Te$_3$/MnBi$_2$Te$_4$ with $x$=0 is within the bulk conduction band regardless of the spacer-thickness [36,39], an RKKY-like ferromagnetic interaction can be mediated by electrons in the bulk conduction band [30,72]. In brief, our present study provides the following scenario: while the ferromagnetic order of the $N$ = 0 system can be described by the bulk-carrier-mediated mechanism, that of the $N \geq 2$ systems is enhanced by the appearance of the van Vleck mechanism due to the bulk band inversion. If we consider the case of Bi$_2$Te$_3$-based systems where the $E_F$ position is within the conduction band, we can exclude the BR mechanism which requires $E_F$ to lie in the band gap [60,61].

On the other hand, in the MBST/BST/MBST sandwich structure where $E_F$ is located within the bulk gap, the top-gate-voltage-dependence as shown in Fig. 9 shows that the ferromagnetic characteristic (AHE coercivity) is scarcely dependent on the $E_F$ position. As mentioned in Section III D, when ferromagnetism is induced by the carrier-independent interaction, the AHE coercivity has been reported to be robust against electrical gating [52]. In addition, as illustrated in Fig. 10, $T_C$ of our MBST systems is larger as the in-plane Mn-Mn distance is reduced, which is consistent with the behavior by the BR mechanism [60]. Thus, the ferromagnetism in our systems can be explained by the carrier-independent mechanism such as the BR model [60,61,63,64] or the hybrid of BR and van Vleck models [4,52,66-69].

In 2022, we previously reported that in the MBST/BST/MBST sandwich structure, nontrivial magnetic vortices, skyrmions, were observed with a proper BST spacer-thickness (1 QL) and with $E_F$ tuned around the gapped surface state [14]. The skyrmions' stability there depends on the $E_F$ position and they are most stabilized as $E_F$ is near CNP. Generally, skyrmions are induced by the competition between ferromagnetic interaction and Dzyaloshinskii-Moriya interaction (DMI). When considered in conjunction with the results of the present study, it is found that the ferromagnetic interaction in this sandwich structure showing skyrmions is not carrier-mediated whereas DMI can be mediated by the Dirac carriers [57].

V. CONCLUSION

In summary, we systematically investigate the inter- and intralayer magnetic interactions in the MBST/BST heterostructure and the MBST/BST/MBST sandwich structure by XRD, magnetic and Hall effect measurements. Through the comparison of the magnetic properties between the heterostructure and the sandwich structure, it is found that the interlayer interaction between the top and bottom MBST layer is ferromagnetic. Then, the dependence of magnetic properties on the BST spacer thickness of the sandwich structure indicates that the two MBST-SLs are ferromagnetically coupled even at a distance of 10 nm. Furthermore, the Sb-content ($x$)-dependence of $T_C$ demonstrates that the intralayer ferromagnetic interaction becomes stronger by decreasing the in-plane lattice constant which corresponds to the nearest Mn-Mn distance within each SL. The mechanism of these magnetic interactions cannot be explained only by the direct exchange and superexchange interactions which were previously reported in MBST systems [20,21,33]. In particular, $T_C$ of the MnBi$_2$Te$_4$/Bi$_2$Te$_3$/MnBi$_2$Te$_4$ (without Sb) sandwich structure where $E_F$ is located in the bulk conduction band with spacer ($N \geq 2$) is larger than that without the spacer ($N = 0$). This indicates that in the case of $N \geq 2$, the van Vleck mechanism related to the nontrivial band topology is effective and then enhances the ferromagnetic order. These interpretations are consistent with an experiment observing ultrafast magnetic dynamics saying that even in a $MnBi_2Te_4$ bulk crystal the ferromagnetism can be induced by the nontrivial band structure rather than by the direct/superexchange interactions [41]. Moreover, the independence of AHE on the top-gate voltage in the MBST/BST/MBST (with Sb-content, $x = 0.55$) sandwich structure suggests that the carrier-independent ferromagnetic mechanism such as the Bloembergen-Rowland and/or van Vleck models works at least when $E_F$ lies within the bulk band gap. Our work provides a deep insight into the mechanisms of magnetic order in MBST systems by revealing novel facts of intra- and interlayer magnetic interactions in detail. Our thin film system would give a clue for increasing the temperature for appearing quantum phenomena for the development of next-generation spintronics and electronics devices.

**Acknowledgement**

The authors thank T. Hirahara in the Tokyo Institute of Technology, T. T. Sasaki in NIMS, S. Kuroda in University of Tsukuba, and G. Springholz in Johannes Kepler University of Linz for fruitful discussions. H. Takagi and K. Kitagawa in the University of Tokyo are acknowledged for their helps in the transport and the XRD systems. R. Toda and T. Fujii in the University of Tokyo are acknowledged for their helps in the transport and magnetic systems. This research was partly supported by JSPS KAKENHI Grant Nos. 18K18732, 20H00342, 20H02616, 23H00265, 23H04098, 24K21727, and 24K00551, and by JST SPRING Grant No. JPMJSP2108 and JST PRESTO Grant No. JPMJPR2451.

Magnetic Exchange Interaction in Topological Insulator Thin Films. Nano Lett. **23**, 2483 (2023).

**Figures and Figure Captions**

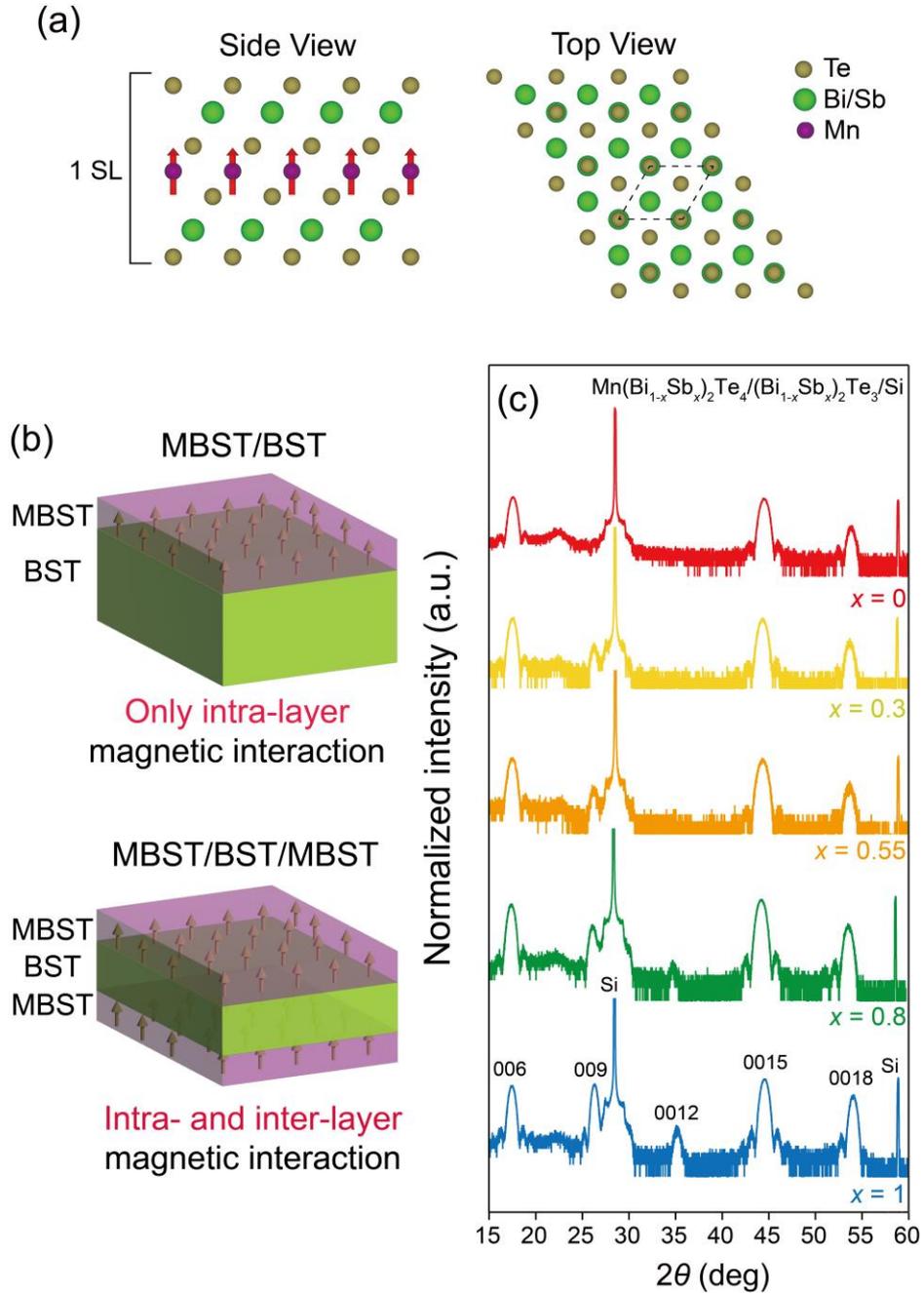

Fig. 1 Schematics of (a) the stacking structure of a septuple layer of MBST, and (b) the heterostructure and sandwich structure composed of BST and MBST layers, respectively. (c) XRD pattern of MBST/BST ($x$ = 0, 0.3, 0.55, 0.8, and 1) heterostructure films grown on Si(111) substrate. The indices on the bottom spectrum are for BST crystal lattice.

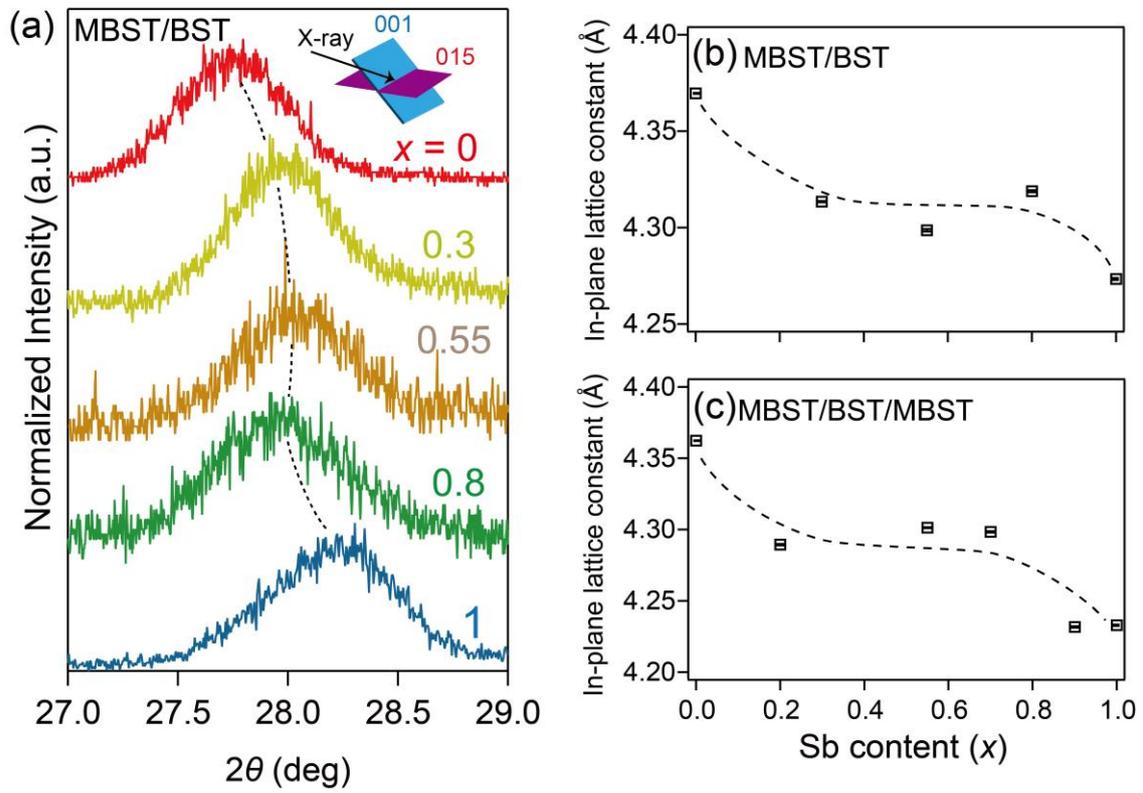

Fig. 2 (a) XRD patterns of MBST/BST with various *x* with respect to the (015) plane of the BST hexagonal crystal, showing peak shift with *x* values. (b,c) Sb-content *x* dependences of the in-plane lattice constants of the MBST/BST and MBST/BST/MBST, respectively.

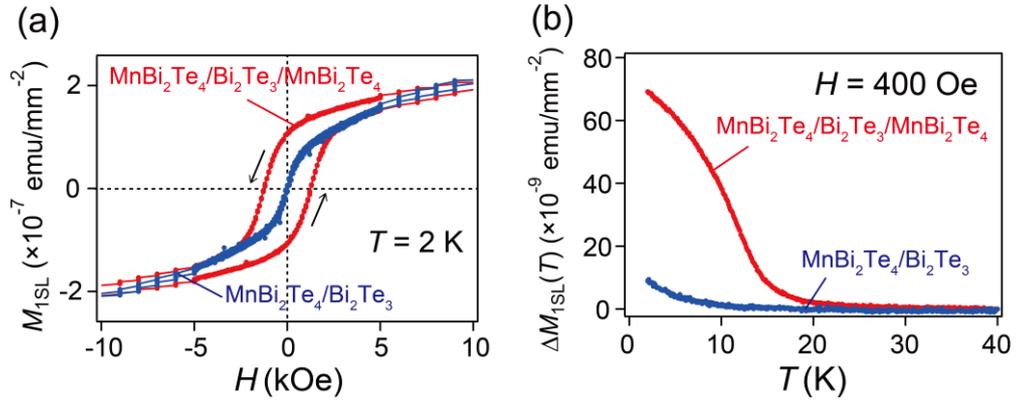

Fig. 3 (a) *M-H* curves under the surface-normal magnetic field in MnBi$_2$Te$_4$/Bi$_2$Te$_3$ and MnBi$_2$Te$_4$/Bi$_2$Te$_3$/MnBi$_2$Te$_4$ at 2 K. $M_{1SL}$ indicates the magnetization normalized for each septuple layer of MnBi$_2$Te$_4$. (b) *M-T* curves under the surface-normal magnetic field (*H* = 400 Oe) in MnBi$_2$Te$_4$/Bi$_2$Te$_3$ and MnBi$_2$Te$_4$/Bi$_2$Te$_3$/MnBi$_2$Te$_4$. Δ$M_{1SL}$ indicates the difference from the value at 40 K, $M_{1SL}$(40 K).

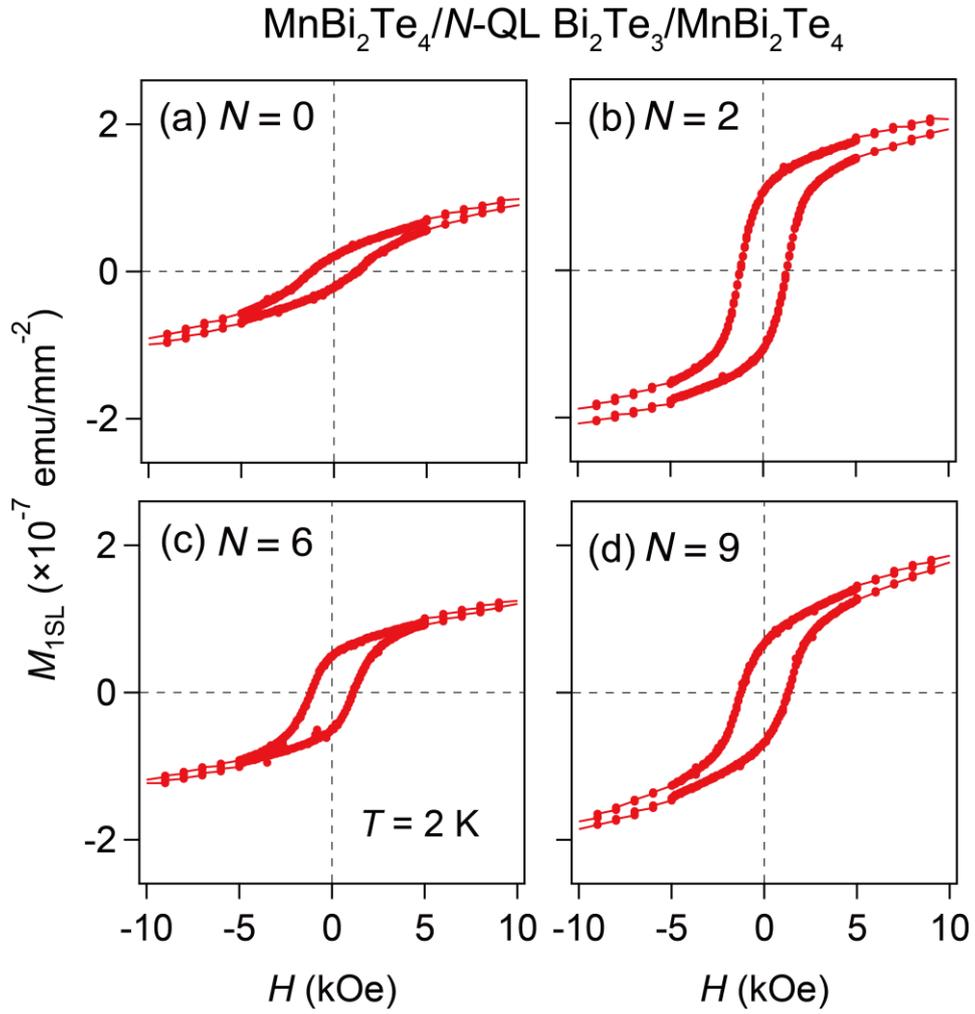

Fig. 4 (a-d) *M-H* curves under the surface-normal magnetic field for $MnBi_2Te_4/Bi_2Te_3/MnBi_2Te_4$ sandwich structures with the spacer $Bi_2Te_3$ layer thickness of 0, 2, 6, and 9 QL measured at 2 K, respectively. $M_{1SL}$ indicates the magnetization normalized for each septuple layer of $MnBi_2Te_4$.

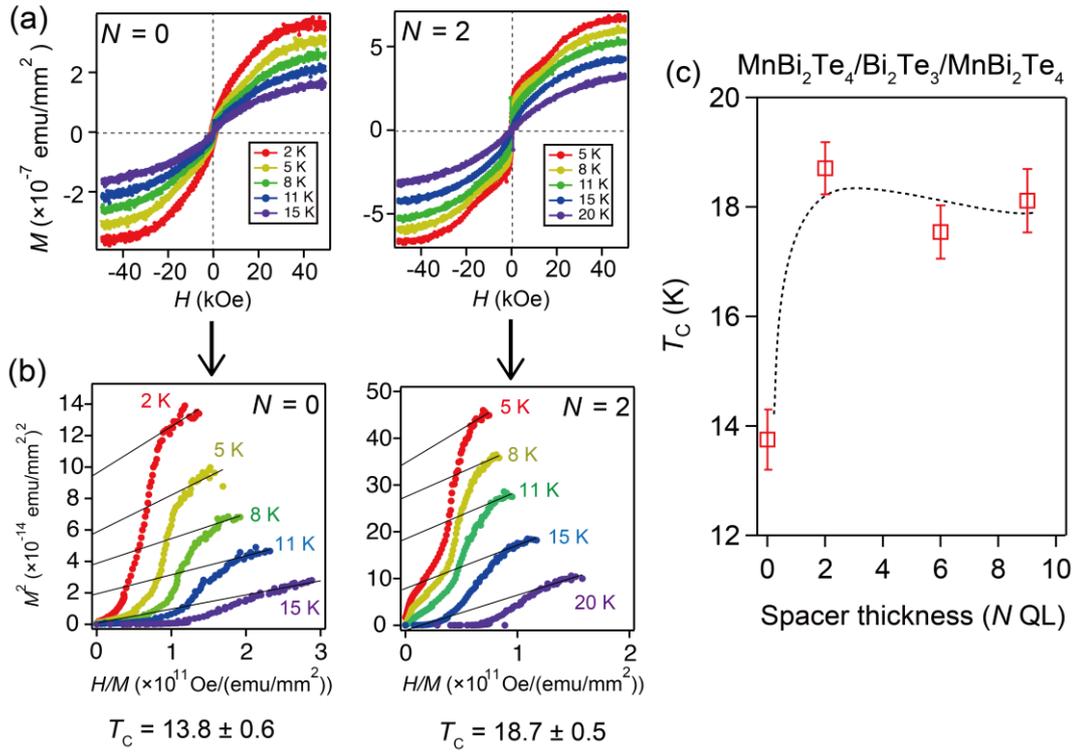

Fig. 5 (a) *M-H* curves under the surface-normal magnetic field at various temperatures in MnBi$_2$Te$_4$/Bi$_2$Te$_3$/MnBi$_2$Te$_4$ sandwich structures with the spacer Bi$_2$Te$_3$ layer thickness of 0, and 2 QL, respectively. (b) Arrott plots from *M-H* curves in (a). (c) The dependence of $T_C$ on the spacer layer (Bi$_2$Te$_3$)-thickness in the sandwich structures at 2 K. Dashed curve is for eye guide.

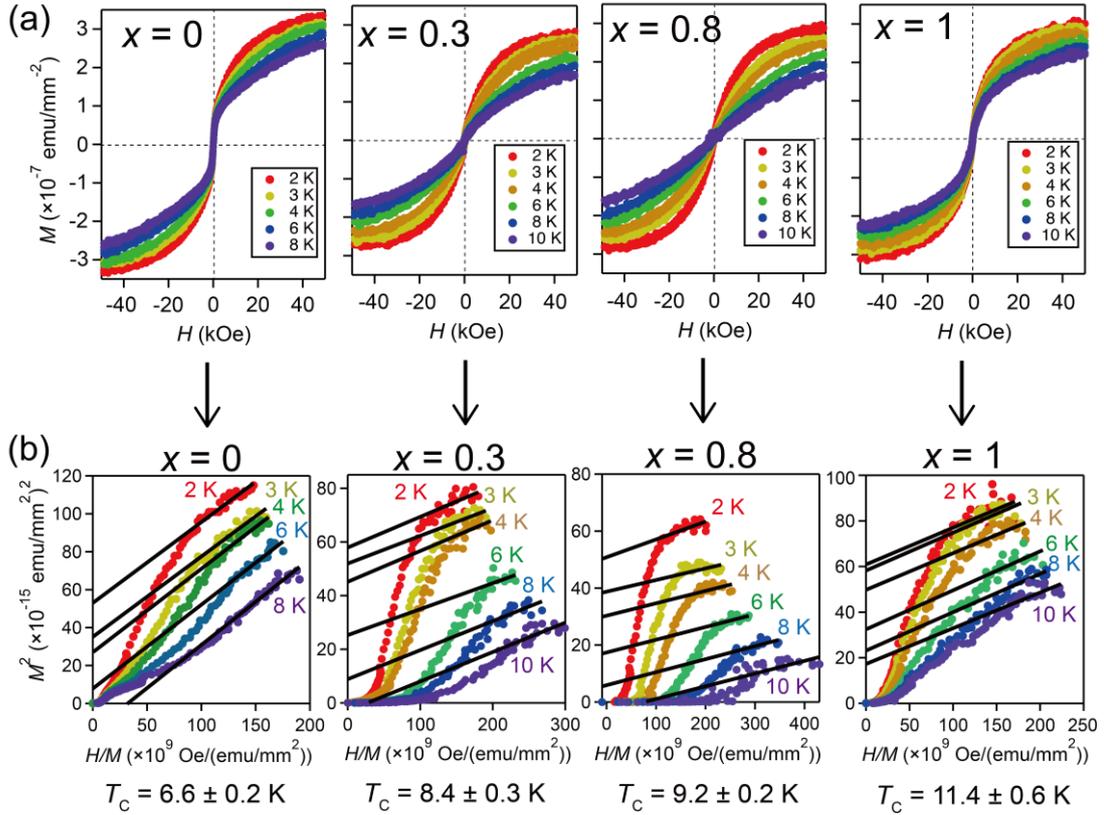

Fig. 6 (a) *M-H* curves under the surface-normal magnetic field at various temperatures in MBST/BST heterostructures with *x* = 0, 0.3, 0.8, and 1. (b) Arrott plots reconstructed from the *M-H* curves in (a).

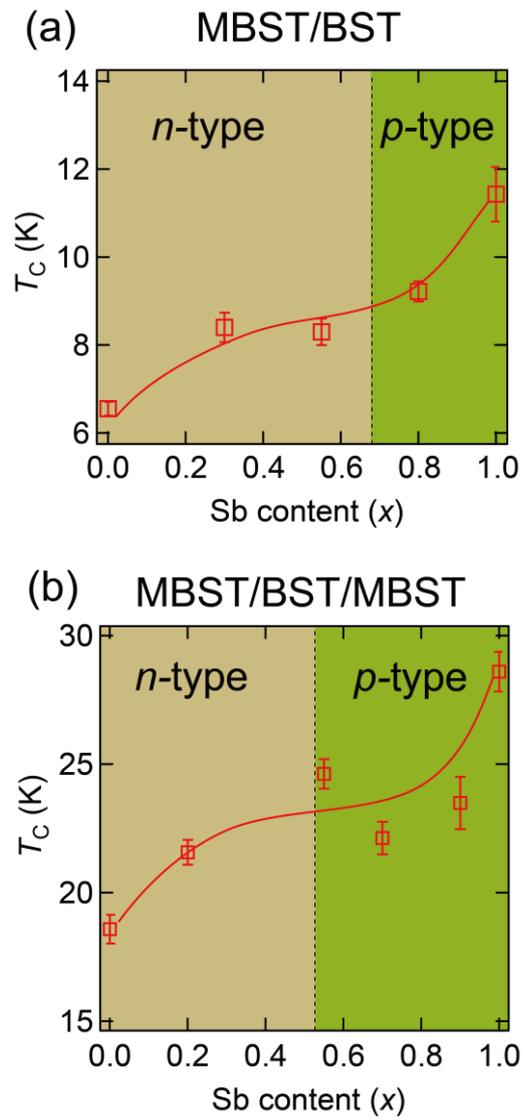

Fig. 7 *x*-dependences of $T_C$ in (a) the MBST/MST heterostructure and (b) the MBST/BST(6 QL)/MBST sandwich structure estimated from Arrott plots in Figs. 5 and 6.

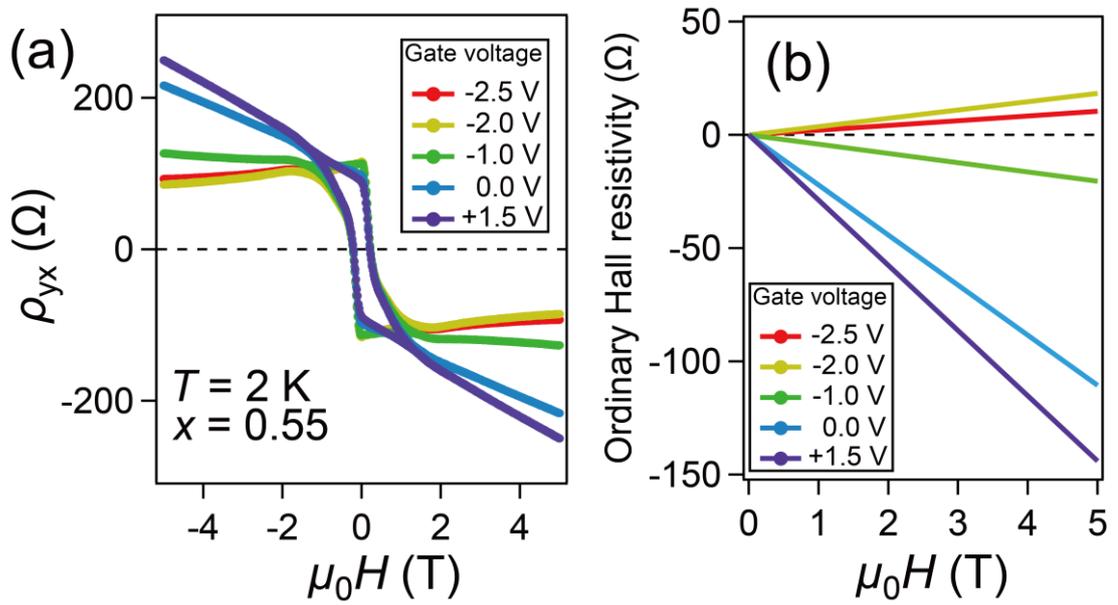

Fig. 8 (a) Magnetic field dependences of Hall resistivity $\rho_{yx}$ of the sandwich structure ($x$ = 0.55) at 2 K at various top gate voltages. The magnetic field is applied normal to the sample surface. (b) Magnetic field dependences of the ordinary Hall resistivity extracted from (a).

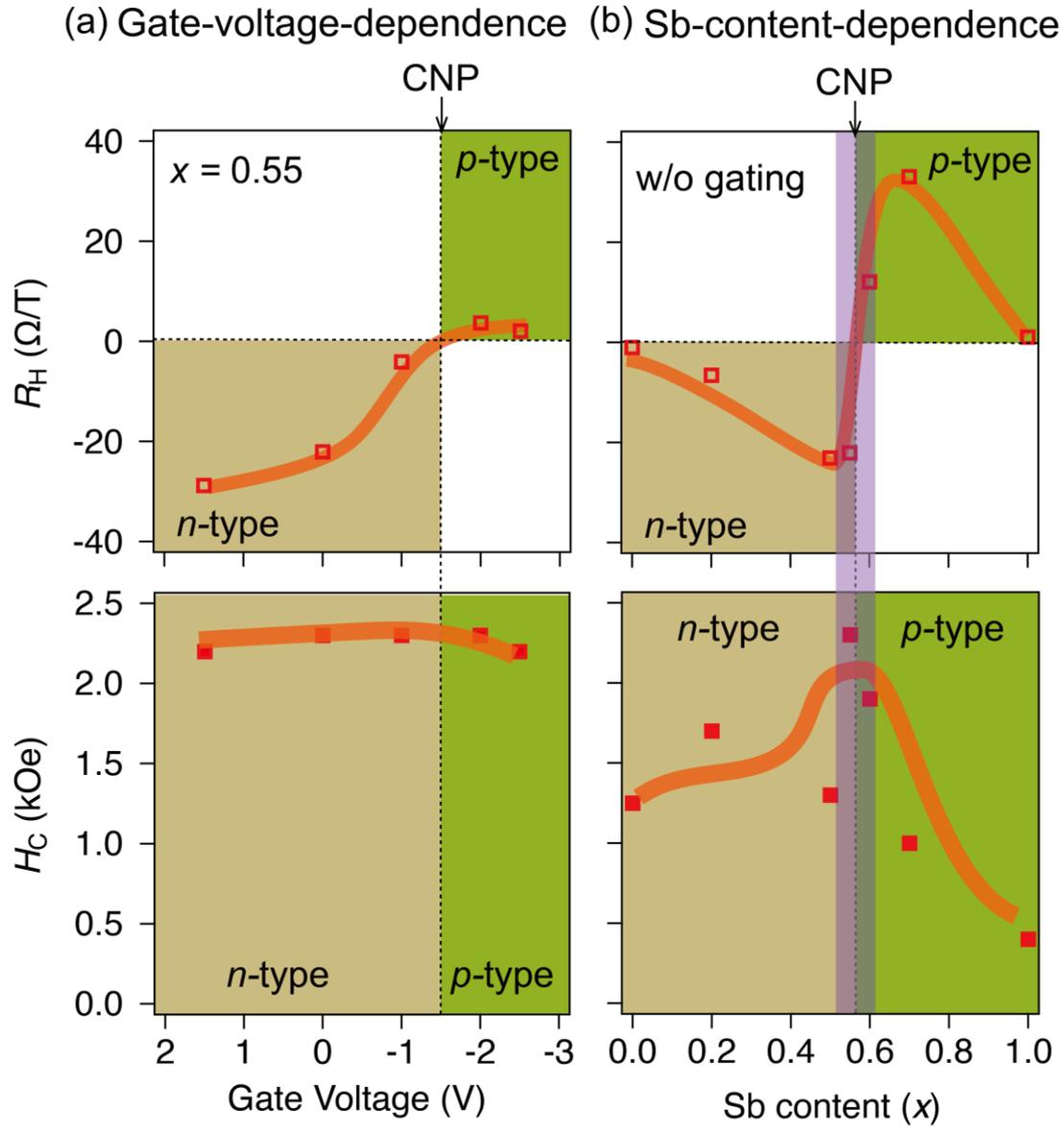

Fig. 9 (a) Top-gate-voltage- and (b) $x$-dependences of the ordinary Hall coefficient $R_H$ (top figures) and the coercivity $H_C$ (bottom figures), derived from the results of Hall effect measurements in the MBST/BST/MBST sandwich structure at 2 K. The curves are for eye guide.

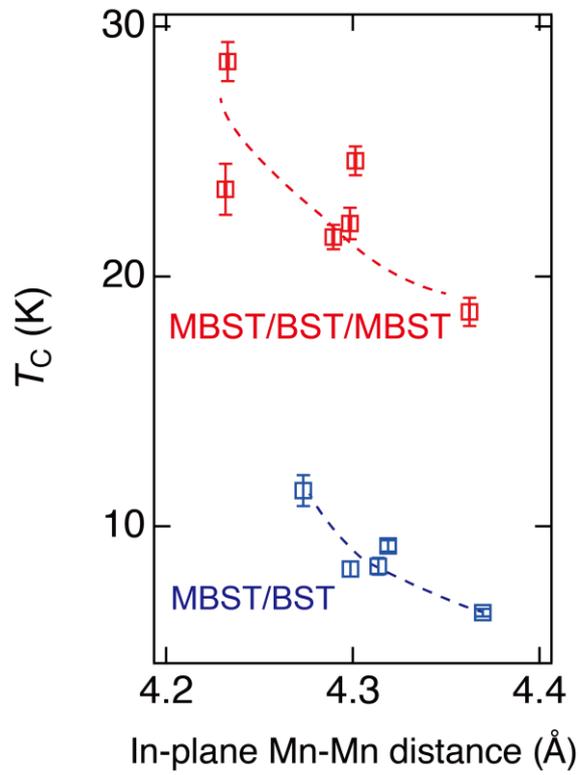

Fig. 10 The dependence of $T_C$ on the in-plane Mn-Mn distance in MBST/BST heterostructures (blue) and MBST/BST/MBST sandwich structures (red). $T_C$ is estimated by Arrott plots as in Fig. 7 and the in-plane Mn-Mn distance is estimated from XRD in Fig. 2.